\documentstyle[12pt]{article}
\begin{document}
\begin{titlepage}

\title{On the gravitational angular momentum of rotating sources}

\author{J. W. Maluf$\,^{*}$ and S. C. Ulhoa\\
Instituto de F\'{\i}sica, \\
Universidade de Bras\'{\i}lia\\
C. P. 04385 \\
70.919-970 Bras\'{\i}lia DF, Brazil\\}
\date{}
\maketitle

\begin{abstract}
The gravitational energy-momentum and angular momentum satisfy
the algebra of the Poincar\'e group in the full phase space of the 
teleparallel equivalent of general relativity. The expression for 
the gravitational energy-momentum may be written as a surface 
integral in the three-dimensional spacelike hypersurface, whereas 
the definition for the angular momentum is given by a volume 
integral. It turns out that in practical calculations of the 
angular momentum of the gravitational field generated by 
localized sources
like rotating neutron stars, the volume integral reduces to a 
surface integral, and the calculations can be easily carried out. 
Similar to previous investigations in the literature, we
show that the total angular momentum is finite provided a certain
asymptotic behaviour is verified. We discuss the dependence of
the gravitational angular momentum on the frame, and argue that it
is a measure of the dragging of inertial frames.
\end{abstract}
\thispagestyle{empty}
\vfill
\par
\bigskip
\noindent (*) e-mail: wadih@unb.br\par
\end{titlepage}
\newpage

\noindent

\section{Introduction}
The teleparallel equivalent of general relativity (TEGR) is a
consistent geometrical framework for the dynamics of the gravitational 
field. The theory is formulated in terms of tetrad fields. It is known 
that these fields are necessary for establishing the interaction 
of Dirac spinor fields with the gravitational field, as well as for 
the description of reference frames in spacetime. In the general case, 
i.e., if no boundary conditions are imposed on the field quantities, 
the TEGR is invariant under global SO(3,1) frame transformations.  
Therefore six degrees of freedom of the tetrad field (with respect to 
the metric tensor) cannot be removed by gauge transformations as in 
the Einstein-Cartan theory, for instance, which exhibits local SO(3,1) 
symmetry. On the other hand, the reference frame is characterized by 
the six components of the acceleration tensor \cite{Maluf1}, which
characterizes the inertial properties of the frame. It follows
that in the TEGR the tetrad field determines both the reference frame 
and the gravitational field. 

The field equations of the theory (Euler-Lagrange and first class
constraint equations) are interpreted as equations that define the 
energy, momentum and angular momentum of the gravitational field. The
interpretation of a constraint equation as an energy equation for a
physical system is not a specific feature of the TEGR. It takes place, 
for instance, in the consideration of Jacobi's action \cite{Lanczos,HG}
for a parametrized nonrelativistic particle. In order to understand
this feature, let us consider a particle of mass $m$ described in the
configuration space with generalized coordinates $q^i$, 
$i=1,2,3$. The particle is subject to the potential $V(q)$ and has 
constant energy $E$. Denoting $\dot{q}^i=dq^i/dt$, where $t$ is a
monotonically increasing parameter between the (fixed) initial and end 
points of the path, the Jacobi action integral for this particle can 
be written as \cite{BY1}

$$I=\int_{t_1}^{t_2} dt
\sqrt{m g_{ij}(q) \dot{q}^i \dot{q}^j}
\sqrt{2\lbrack E-V(q)\rbrack}\,.$$

\noindent The action is extemized by varying the configuration space
path and requiring $\delta q(t_1)=\delta q(t_2)=0$. We may simplify 
the integrand by writing $dt \sqrt{m g_{ij}\dot{q}^i \dot{q}^j}=
\sqrt{m g_{ij} dq^i dq^j}$, which shows that the action is invariant
under reparametrizations of the time parameter $t$. 
Thus in Jacobi's formulation of the action principle it is
the energy $E$ of the particle that is fixed, not its initial and
final instants of time. In view of the time reparametrization of the
action integral, the Hamiltonian constructed out of the Lagrangian 
above vanishes identically, which is a known feature of 
reparametrization invariant theories. 
If we denote $p_i$ as the momenta conjugate to $q^i$, we find
$p_i=g_{ij} \dot{q}^j\sqrt{2(E-V)/\dot{q}^2}$ (where 
$\dot{q}^2=g_{kl}\dot{q}^k\dot{q}^l$), which leads to the 
constraint

$$C(q,p)\equiv {1\over 2}g^{ij}p_i p_j + V(q)-E\approx 0$$

\noindent The equation of motion obtained from the action integral
has to be supplemented by the constraint equation $C=0$, in order to 
be equivalent with Newton's equation of motion with fixed energy $E$ 
\cite{BY1}. Therefore we see that the constraint equation defines
the energy of the particle. This is the feature that takes place in
the TEGR: the definitions of the energy-momentum and angular momentum 
of the gravitational field emerge from the constraint equations of the 
theory \cite{Maluf2,Maluf3}. These definitions are viable as long as
they yield consistent values in the consideration of well understood
gravitational field configurations.

The evaluation of the expression for the gravitational energy-momentum
in the TEGR is not difficult because it reduces to a surface integral
in the three-dimensional spacelike hypersurface. 
However, the gravitational angular momentum is given by a volume 
integral, and we did not succeed in reducing the latter to a surface 
integral in the general case. Surprisingly, we find that in the 
framework of static observers the expression for the gravitational 
angular momentum due to the most general rotating source reduces to a 
simple surface integral. This feature
not only allows the evaluation of the expression in an easy way, but 
it also points out which is the necessary fall-off of the components 
of the metric tensor that yields a finite value for the total
angular momentum. We find that if the metric tensor obeys 
asymptotically flat boundary conditions, the total gravitational 
angular momentum will be finite provided the metric components  
fall off as $1/r$ or faster in the limit $r \rightarrow \infty$. This
result agrees with the detailed analysis of Beig and \'O Murchadha 
\cite{BM} and of Szabados \cite{Sza}, who investigated the Poincar\'e
structure of asymptotically flat spacetimes.

In this paper we analyze in detail the gravitational angular 
momentum of a pulsar model that describes a rotating neutron star. If
$J$ represents the angular momentum of the source, we find that the
gravitational angular momentum is given by $2J/3$. According to our
analysis this result seems to be general: if $J$ is the angular
momentum of an arbitrary rotating source, the gravitational 
angular momentum is given by $2J/3$. 
The constraint equation that defines the gravitational angular 
momentum is not related to the energy-momentum tensor. Thus we 
interpret the angular momentum discussed here as the angular momentum
of the field, not of the source. 

In the Newtonian description of classical
mechanics the angular momentum of the source is frame dependent. This
feature also holds in relativistic mechanics. If the angular momentum
of the source in general is frame dependent, in principle one cannot
exclude the possibility that the angular momentum of the field is 
frame dependent as well. 
Differently from other definitions of gravitational angular momentum 
that are formulated in terms of surface integrals at spacelike
infinity and depend only on the asymptotic behaviour of the
metric tensor, the definition considered here naturally depends 
on the frame, as it is invariant under global 
(rather than local) SO(3,1) transformations of the tetrad field.
In the present framework, 
observers that are in rotational motion around the rotating source
measure the gravitational angular momentum differently from 
static observers. These observers also measure 
the angular momentum of the source differently from stationary
observers (in Newtonian mechanics the angular 
momentum of the source in the frame of observers that 
rotate at the same angular frequency vanishes). We will show that
the concept of the present definition of gravitational angular 
momentum is closely related to the dragging of inertial frames. 

In section 2 we present a summary of the formulation of the TEGR.
In section 3 we address the gravitational angular momentum of a 
rotating neutron star, and then of an arbitrary gravitational field
configuration. In section 4 we argue that the dependence of the 
gravitational angular momentum on the frame is a natural feature of
the present formalism. Observers whose angular velocity around the 
rotating source is the same as the dragging velocity do not measure 
dragging effects (the dragging velocity, for instance)
as do static observers. We show that for such
observers the gravitational angular momentum vanishes.
Finally, in section 5 we present the final remarks.

Notation: space-time indices $\mu, \nu, ...$ and SO(3,1)
indices $a, b, ...$ run from 0 to 3. Latin indices from the middle
of the alphabet $(i,j,k)$ run from 1 to 3. Time and space indices are
indicated according to
$\mu=0,i,\;\;a=(0),(i)$. The tetrad field is denoted by $e^a\,_\mu$,
and the torsion tensor reads
$T_{a\mu\nu}=\partial_\mu e_{a\nu}-\partial_\nu e_{a\mu}$.
($T_{a\mu\nu}$ is related
via $T_{a\mu\nu}=e_{a \lambda} T^\lambda\,_{\mu\nu}$ to the torsion
tensor of Cartan's connection 
$\Gamma^\lambda_{\mu\nu}=e^{a\lambda}\partial_\mu e_{a\nu}$).
The flat, Minkowski space-time metric tensor raises and lowers
tetrad indices and is fixed by
$\eta_{ab}=e_{a\mu} e_{b\nu}g^{\mu\nu}= (-+++)$. The determinant of 
the tetrad field is represented by $e=\det(e^a\,_\mu)$.\par        
\bigskip

\section{The Hamiltonian form of the TEGR}

The equivalence of the TEGR with Einstein's general relativity may be
understood by means of an identity between the scalar curvature 
$R(e)$ constructed out of the tetrad field and a combination of
quadratic terms of the torsion tensor (see, for example, 
ref. \cite{Maluf4}), 

\begin{equation}
eR(e)\equiv -e({1\over 4}T^{abc}T_{abc}+{1\over 2}T^{abc}T_{bac}-T^aT_a)
+2\partial_\mu(eT^\mu)\,.
\label{1}
\end{equation}

The Lagrangian density of the TEGR in empty spacetime is given by 
the combination of the quadratic terms on the right hand side of 
eq. (1),

\begin{eqnarray}
L&=& -k e({1\over 4}T^{abc}T_{abc}+{1\over 2}T^{abc}T_{bac}-
T^aT_a) \nonumber \\
&\equiv& -ke\Sigma^{abc}T_{abc}\,, 
\label{2}
\end{eqnarray}
where $k=c^3/16\pi G$, $T_a=T^b\,_{ba}$, 
$T_{abc}=e_b\,^\mu e_c\,^\nu T_{a\mu\nu}$ and

\begin{equation}
\Sigma^{abc}={1\over 4} (T^{abc}+T^{bac}-T^{cab})
+{1\over 2}( \eta^{ac}T^b-\eta^{ab}T^c)\;.
\label{3}
\end{equation}
The field equations derived from eq. (2) are 
equivalent to Einstein's equations.

The Hamiltonian formulation of the theory has been investigated 
in detail in ref. \cite{Maluf5}. The Hamiltonian density $H$ may be
obtained from $L$ by rewriting it as $L=p\dot{q}-H$, and expressing
$H$ in terms of the canonical variables. There is no time derivative
of $e_{a0}$, and therefore this field quantity arises as a
Lagrange multiplier. The momenta canonically conjugated to $e_{ak}$
is denoted by $\Pi^{ak}$. In the configuration space we have

\begin{equation}
\Pi^{ai}=-4ke\Sigma^{a0i}\,.
\label{4}
\end{equation}
We find that $H$ may indeed be expressed 
in terms of $e_{ak}$, $\Pi^{ak}$ and Lagrange multipliers. In ref.
\cite{Maluf3} we have redefined some Lagrange multipliers and 
constraints in order to arrive at simpler form of $H$. Except for
a surface term it reads

\begin{equation}
H=e_{a0}C^a+{1\over 2}\lambda_{ab}\Gamma^{ab}\,,
\label{5}
\end{equation}
where $e_{a0}$ and $\lambda_{ab}=-\lambda_{ba}$ 
are Lagrange multipliers, and $C^a$ and $\Gamma^{ab}$ are 
first class constraints. After solving one set of Hamilton's
field equations we may identify $\lambda_{ik}=1/2(T_{i0k}-T_{k0i})$
and $\lambda_{0k}=T_{00k}$. The quantities $\lambda_{ik}$ and
$\lambda_{0k}$ are components of 
$\lambda_{\mu\nu}=e^a\,_\mu e^b\,_\nu \lambda_{ab}$.
For further details see refs. \cite{Maluf3,Maluf5}.

The constraint $C^a$ may be written in the form 
$C^a=-\partial_i\Pi^{ai}+h^a$, where $h^a$ is an intricate 
expression of the field quantities. We note that 
$-\partial_i\Pi^{ai}$ is the only total divergence of the momenta
$\Pi^{ai}$ that arises in the expression of $C^a$. The integral form
of the equation $C^a=0$ motivates the definition of the gravitational
energy-momentum $P^a$ \cite{Maluf2},

\begin{equation}
P^a=-\int_V d^3x\,\partial_i \Pi^{ai}\,,
\label{6}
\end{equation}
where $V$ is an arbitrary volume of the three-dimensional space.
The emergence of
a nontrivial total divergence is a feature of theories with torsion.
Metric theories of gravity do not share this feature. The 
integration of this total divergence yields a surface integral. If we 
consider the $a=(0)$ component of eq. (6) and adopt asymptotic 
boundary conditions for the tetrad field we find \cite{Maluf2} that 
the resulting 
expression is precisely the surface integral at infinity that
defines the ADM energy \cite{ADM}. This fact is a strong indication 
(but no proof) that eq. (6) does indeed represent the gravitational
energy-momentum.

The constraint $\Gamma^{ab}$ is defined by \cite{Maluf3,Maluf5}

\begin{equation}
\Gamma^{ab}=M^{ab}+4ke(\Sigma^{a0b}-\Sigma^{b0a})\,,
\label{7}
\end{equation}
where $M^{ab}=e^a\,_\mu e^b\,_\nu M^{\mu\nu}=-M^{ba}$, and 
$M^{\mu\nu}$ is given by

\begin{eqnarray}
M^{ik}&=&2\Pi^{\lbrack ik \rbrack}=e_a\,^i \Pi^{ak}-
e_a\,^k \Pi^{ai}\,, \\
M^{0k}&=&\Pi^{0k}=e_a\,^0 \Pi^{ak}\,. 
\label{8,9}
\end{eqnarray}

Similar to the definition of $P^a$, the integral form of 
the constraint equation $\Gamma^{ab}=0$ motivates the definition
of the space-time angular momentum. The equation $\Gamma^{ab}=0$
implies

\begin{equation}
M^{ab}=-4ke(\Sigma^{a0b}-\Sigma^{b0a})\,. 
\label{10}
\end{equation}
Therefore we define \cite{Maluf3}

\begin{equation}
L^{ab}=-\int_V d^3x\; e^a\,_\mu e^b\,_\nu M^{\mu\nu}\,,
\label{11}
\end{equation}
as the 4-angular momentum of the gravitational field. This 
definition differs by a sign from the definition presented in
ref. \cite{Maluf3}.

The evaluation of definitions (6) and (11) are carried out in the
configuration space, by means of eqs. (4) and (10), respectively.
However, if we consider (6) and (11) in the phase space of the 
theory, with the right hand side of the latter equations given in
terms of the momenta $\Pi^{ai}$ (with the help of eqs. (8) and (9)),
we find that $P^a$ and $L^{ab}$ constitute a representation of the
Poincar\'{e} group. By working out the Poisson brackets between 
$P^a$ and $L^{ab}$ in the full phase space of the theory we find
\cite{Maluf3}

\begin{eqnarray}
\lbrace P^a , P^b \rbrace &=& 0\,, \nonumber \\
\lbrace P^a , L^{bc} \rbrace &=& \eta^{ab} P^c -\eta^{ac} P^b 
\,, \nonumber \\
\lbrace L^{ab}, L^{cd} \rbrace &=&
\eta^{ac}L^{bd} +\eta^{bd}L^{ac} -\eta^{ad}L^{bc}-\eta^{bc}L^{ad}
\,.
\label{12}
\end{eqnarray}
The Poincar\'{e} algebra of $P^a$ 
and $L^{ab}$ confirms the consistency of the definitions. We recall 
that the evaluation of $P^a$ and $L^{ab}$ is carried out in an
arbitrary  volume $V$ of the three-dimensional space. 

Two important
points must be noted. First, the Poincar\'e
algebra above is rigorously verified irrespective of the 
specification of any surface term (surface integral) on the boundary 
of $V$. Thus it is not necessary to impose boundary conditions to
obtain (12). Second, the procedure for obtaining (12) is {\it exactly
the same} procedure for obtaining the algebra of constraints of a 
theory in Hamiltonian form (as in ref. \cite{Maluf5}, for instance).
Therefore the validity of the constraint algebra and of eq. (12) are
on equal footing.

Definitions (6) and (11) are invariant under (i) general coordinate
transformations of the three-dimensional space, (ii) time 
reparametrizations, and (iii) global SO(3,1) transformations. The
non-invariance of these definitions under the local SO(3,1) group
reflects the frame dependence of the definitions. We have argued
\cite{Maluf1,Maluf3,GRG} that this dependence is a natural feature 
of the gravitational energy-momentum defined by eq. (6).
In the TEGR each set of tetrad
fields is interpreted as a reference frame in spacetime. The 
tetrad frame is adapted to observers whose four-velocity $u^\mu$ in 
spacetime is identified with $e_{(0)}\,^\mu$, namely, 
$e_{(0)}\,^\mu=u^\mu$, and is characterized by the acceleration 
tensor $\phi_{ab}$ defined by \cite{Maluf1}

\begin{equation}
\phi_{ab}={1\over 2} \lbrack T_{(0)ab}+T_{a(0)b}-T_{b(0)a}
\rbrack\,.
\label{13}
\end{equation}
The tensor $\phi_{ab}$ characterizes the inertial properties of the
frame. It is clearly not invariant under the local SO(3,1) group. It
provides the values of the inertial accelerations that are necessary
to maintain the frame in a given orientation and inertial state, in 
the gravitational field configuration defined by $g_{\mu\nu}$. 

\section{The gravitational angular momentum of rotating sources}

\subsection{A simple model for rotating neutron stars}

The starting point of the present analysis is the analytic pulsar
model described in ref. \cite{AJ1}, which is taken to represent a
rotating neutron star. The model represents a star that
is approximately rigidly rotating.
In the latter reference the authors argue that although the stellar
rotation rate is a function of radius, it can be made arbitrarily
close to rigid rotation. For our purposes the model is important
because it is simple and singularity free, and allows to carry out 
integration over the whole three-dimensional space. It is represented
by the metric tensor

\begin{equation}
ds^2=-\alpha^2dt^2+\beta^2dr^2+r^2d\theta^2+r^2\sin^2\theta
(d\phi-\Omega dt)^2\,.
\label{14}
\end{equation}

We denote by $R$ the stellar radius. For $r \le R$ we have

\begin{eqnarray}
\alpha&=&{1\over 2}\lbrack 3(1-8\pi\rho R^2/3)^{1/2}-
(1-8\pi\rho r^2/3)^{1/2}\rbrack \,, \nonumber \\
\beta^2&=&(1-8\pi\rho r^2/3)^{-1}\,, \nonumber \\
\Omega&=&\Omega(0)\lbrack 1- b(r/R)^2-b\tau (r/R)^4\rbrack \,, 
\label{15}
\end{eqnarray}
and for $r \ge R$,

\begin{eqnarray}
\alpha^2&=& \lbrack 1-2m(R)/r \rbrack\,, \nonumber \\
\beta^2&=& \lbrack 1-2m(R)/r \rbrack^{-1}\,, \nonumber \\
\Omega&=&{{2J}\over r^3}={{2GJ_s}\over c^3r^3} \,,
\label{16}
\end{eqnarray}
where $\Omega$ is the angular velocity of inertial frames along the
rotation axis and $J_s$ represents the angular momentum of the star;
$\rho$ is the uniform density and $m(r)=4\pi\rho r^3/3$. The 
quantity $b$ is defined by $b=3/(5+7\tau)$, where $\tau$ is a free
parameter. The case $\tau=0$ yields completely rigid rotation in the
Newtonian limit. For further details, see ref. \cite{AJ1}.

The evaluation of definition (11) is conceptually simple. It 
requires the calculation of the tensors $T_{\lambda\mu\nu}$ and
$\Sigma^{\lambda\mu\nu}$. We only need to determine the tetrad field
adapted to a particular class of observers. We choose to work with
the tetrad field adapted to stationary observers. Such observers have
the velocity field $u^{\lambda}=(u^0,0,0,0)$. 
Therefore we identify $e_{(0)}\,^\mu=u^\mu $ and require 

\begin{equation}
e_{(0)}\,^i=0\,.
\label{17}
\end{equation}
Moreover we choose the $e_{(3)}\,^\mu$ component to coincide
asymptotically ($r\rightarrow \infty$) with the unit vector
$\hat{\bf z}$ along the $z$ axis, namely,

\begin{equation}
e_{(3)}\,^\mu (t,x,y,z)
\cong (0,0,0,1)\,.
\label{18}
\end{equation}
These two conditions were imposed 
on the set of tetrad fields for the Kerr spacetime, discussed in
ref. \cite{Maluf1}. They do not fix the tetrad field completely
because of the axial symmetry of the spacetime. The orientations of
the $x$ and $y$ directions are arbitrary, except that the latter are
orthogonal. 

The requirement of
the conditions above on $e_{a\mu}$ constructed out of (14) yields

\begin{equation}
e_{a\mu}(t,r,\theta,\phi)=\pmatrix{-A&0&0&-C\cr
0&B \,\sin\theta \cos\phi&
r \cos\theta \cos\phi & -D\, r \sin\theta \sin\phi\cr
0& B\, \sin\theta \sin\phi&
r \cos\theta \sin\phi &  D\, r \sin\theta \cos\phi\cr
0& B\, \cos\theta & -r\sin\theta&0}\,,
\label{19}
\end{equation}
with the following definitions,

\begin{eqnarray}
A(r,\theta)&=& (-g_{00})^{1/2}\,, \nonumber \\
B(r)&=&(g_{11})^{1/2}\,, \nonumber \\
C(r,\theta)&=&-{{g_{03}}\over{(-g_{00})^{1/2}}}\,, \nonumber \\
D(r,\theta)
&=&\biggl( {{-\delta} \over {g_{00}g_{33}}}
\biggr)^{1/2}\,, \nonumber \\
\delta&=& g_{03}g_{03}-g_{00}g_{33}\,.
\label{20}
\end{eqnarray}

The gravitational angular momentum density along the $z$ direction
is given by $M^{(1)(2)}$. Taking into account the definition of
$\Sigma^{abc}$ in (10) we find

\begin{equation}
M^{(1)(2)}=-2ke\,e^{(1)}\,_\mu e^{(2)}\,_\nu
\lbrack T^{0\mu\nu}-g^{0\mu}T^\nu+g^{0\nu}T^\mu \rbrack\,.
\label{21}
\end{equation}
The determinant $e$ reads  
$e=r(g_{11}\,\delta)^{1/2}$. We recall that
$T^\mu=T^\lambda\,_\lambda\,^\mu$. The following relations are valid
for the metric tensor in spherical coordinates for the spacetime of
an arbitrary rotating source,

\begin{eqnarray}
g^{00}&=&- {g_{33} \over {\delta}}\,, \nonumber \\
g^{03}&=&{g_{03} \over {\delta}}\,, \nonumber \\
g^{33}&=& - {g_{00} \over {\delta}}\,.
\label{22}
\end{eqnarray}

In view of (19) we have

\begin{eqnarray}
e^{(1)}\,_1 e^{(2)}\,_2- e^{(1)}\,_2 e^{(2)}\,_1&=&0 \,,\nonumber \\
e^{(1)}\,_1 e^{(2)}\,_3- e^{(1)}\,_3 e^{(2)}\,_1&=&
BD\,r\sin^2\theta \,, \nonumber \\
e^{(1)}\,_2 e^{(2)}\,_3- e^{(1)}\,_3 e^{(2)}\,_2&=&
D\,r^2\sin\theta\,\cos\theta\,,
\label{23}
\end{eqnarray}
and $M^{(1)(2)}$ is first simplified as

\begin{eqnarray}
M^{(1)(2)}&=& -2ke\,BD\,r \sin^2\theta 
(T^{013}+g^{03}T^1) \nonumber \\
{}&{}&-2ke\,D\,r^2\sin\theta \cos\theta
(T^{023}+g^{03}T^2) \nonumber \\
&=& -2ke\,BD\,r \sin^2\theta
\lbrack g^{11}(g^{00}g^{33}-g^{03}g^{03})T_{013} \nonumber \\
&{}&- g^{03}g^{11}g^{22}T_{212}\rbrack \nonumber \\
&{}&-2ke\,D\,r^2\sin\theta \cos\theta
\lbrack g^{22}(g^{00}g^{33}-g^{03}g^{03})T_{023} \nonumber \\
&{}&+
g^{03}g^{11}g^{22}T_{112}\rbrack \,.
\label{24}
\end{eqnarray}

The necessary torsion tensor components are 

\begin{eqnarray}
T_{013}&=&-A\partial_1 C\,,\nonumber \\
T_{023}&=&-A\partial_2 C\,,\nonumber \\
T_{112}&=& 0 \,, \nonumber \\
T_{212}&=&r(1-B) \,,
\label{25}
\end{eqnarray}
where $\partial_1=\partial_r$ and $\partial_2=\partial_\theta$.
Considering $g^{00}g^{33}-g^{03}g^{03}=-1/\delta$ we find that 
eq. (24) is written as

\begin{eqnarray}
M^{(1)(2)}&=&-2ke(BDr\sin^2\theta)\biggl[
{g^{11}\over \delta} A \partial_1 C-g^{03}g^{11}g^{22}\,r(1-B)
\biggr] \nonumber \\
&{}&-2ke(Dr^2\sin\theta\cos\theta)\biggl[{g^{22}\over \delta}
A\partial_2 C \biggr]\,.
\label{26}
\end{eqnarray}

After substitution of eq. (20) into (26) we obtain

\begin{eqnarray}
M^{(1)(2)}&=&2kr\sin\theta\;\partial_1 \biggl[
{g_{03} \over {(-g_{00})^{1/2}}}\biggr]
+2k\sin\theta \biggl[{g_{03} \over (-g_{00})^{1/2}}\biggr]
\biggl[1-(g_{11})^{1/2}\biggr] \nonumber \\
&{}&+2k \cos\theta \;(g_{11})^{1/2}\;\partial_2
\biggl[ {g_{03} \over {(-g_{00}^{1/2})}} \biggr]\,,
\label{27}
\end{eqnarray}
which leads to the final form of $M^{(1)(2)}$,

\begin{equation}
M^{(1)(2)}=2k\,{\partial \over {\partial r}}
\biggl[ {{r\,\sin\theta\,g_{03}}\over (-g_{00})^{1/2}}
\biggr]+2k\,{\partial \over {\partial \theta}}
\biggl[ {{\cos\theta\,(g_{11})^{1/2}\,g_{03}}\over (-g_{00})^{1/2}}
\biggr]\,.
\label{28}
\end{equation}

From eq. (14) we have $g_{03}=-\Omega(r)\,r^2\,\sin^2\theta$.
Taking into account definitions (15) and (16) we find

\begin{eqnarray}
\int_0^\infty dr\,
{\partial \over {\partial r}}
\biggl[ {{r\,\sin\theta\,g_{03}}\over (-g_{00})^{1/2}}
\biggr]&=&-2J\sin^3\theta \nonumber \\
\int_0^\pi d\theta\,{\partial \over {\partial \theta}}
\biggl[ {{\cos\theta\,(g_{11})^{1/2}\,g_{03}}\over (-g_{00})^{1/2}}
\biggr]&=&0 \,.
\label{29}
\end{eqnarray}

Therefore we have

\begin{eqnarray}
L^{(1)(2)}&=&-\int d^3x M^{(1)(2)} \nonumber \\
&=&-2k\int_0^{2\pi}d\phi \int_0^\pi 
d\theta(-2J\sin^3\theta)\nonumber \\
&=&{{32\pi}\over 3}k\,J\,.
\label{30}
\end{eqnarray}
Considering that $k=c^3/(16\pi G)$ and $J=(G/c^3)J_s$, where $J_s$ 
is the angular momentum of the star, we finally obtain

\begin{equation}
L^{(1)(2)}={2\over 3} J_s\,.
\label{31}
\end{equation}
The evaluation of $L^{(1)(2)}$ is easy because $M^{(1)(2)}$ can be
expressed as a total divergence, and eventually $L^{(1)(2)}$ is given
by a surface integral. We see that the expression of $L^{(1)(2)}$ is
finite provided $-g_{00}\cong 1+O(1/r)$ and $g_{03}\cong O(1/r)$ at
spacelike infinity. 

The expressions for $L^{(1)(3)}$ and $L^{(2)(3)}$
vanish because $e^{(1)}\,_i e^{(3)}\,_j- e^{(1)}\,_j e^{(3)}\,_i$
and $e^{(2)}\,_i e^{(3)}\,_j- e^{(2)}\,_j e^{(3)}\,_i$ contain the 
functions $\sin\phi$ or $\,\cos\phi$, and the integration of these 
functions from $0$ to $2\pi$ vanishes. We note that the quantities
above arise in the expressions of $M^{(1)(3)}$ and $M^{(2)(3)}$, 
respectively, exactly like  
$e^{(1)}\,_i e^{(2)}\,_j- e^{(1)}\,_j e^{(2)}\,_i$ does in eq. (21).

For the same reason above, we conclude that the 
center-of-mass moments $L^{(0)(1)}$ and $L^{(0)(2)}$ vanish. The 
only component that does not vanish on account of integrals like
$\int_0^{2\pi} d\phi\, \sin\phi=\int_0^{2\pi} d\phi\, \cos\phi=0$
is $L^{(0)(3)}$. From definition (10) we obtain

\begin{eqnarray}
M^{(0)(3)}&=&-2ke\lbrack e^{(0)}\,_0 e^{(3)}\,_1
(T^{001}-g^{00}T^1)\nonumber \\
&{}&+e^{(0)}\,_0 e^{(3)}\,_2
(T^{002}-g^{00}T^2)\nonumber \\
&{}&-e^{(0)}\,_3 e^{(3)}\,_1 (T^{013}-g^{03}T^1)\nonumber \\
&{}&-e^{(0)}\,_3 e^{(3)}\,_2 (T^{023}-g^{03}T^2)\rbrack\,.
\label{32}
\end{eqnarray}
After a long calculation and 
several simplifications we arrive at

\begin{equation}
M^{(0)(3)}=2k\lbrack
\partial_1(D\,r^2\sin\theta\,\cos\theta)-
\partial_2(BD\,r\sin^2\theta)\rbrack \,.
\label{33}
\end{equation}
Since $D$ is a function of $\sin^2\theta$ we conclude that

\begin{equation}
\int_0^\pi d\theta\, \partial_1(D\,r^2\sin\theta\,\cos\theta)=
\int_0^\pi d\theta\, \partial_2(BD\,r\,\sin^2\theta)=0\,,
\label{34}
\end{equation}
and therefore $L^{(0)(3)}=0$. Thus the only nonvanishing component
of $L^{ab}$ is given by eq. (31).

\subsection{An arbitrary rotating source with axial symmetry}

The most general form of the metric tensor that describes the
spacetime generated by an arbitrary stationary
rotating source with axial
symmetry is represented by the line element
(see section 6.4 of ref. \cite{NKG})

\begin{equation}
ds^2=g_{00}dt^2+g_{11}dr^2+g_{22}d\theta^2+g_{33}d\phi^2
+2g_{03}d\phi\, dt\,,
\label{35}
\end{equation}
where {\it all} metric components depend on $r$ and $\theta$:
$g_{\mu\nu}=g_{\mu\nu}(r,\theta)$.
Denoting by $g$ the determinant of the metric tensor, we have
$\sqrt{-g}=\lbrack g_{11}g_{22}
(g_{03}g_{03}-g_{00}g_{33})\rbrack^{1/2}$. The inverse metric 
components $g^{00}$, $g^{03}$ and $g^{33}$ are given by
eq. (22). Here we also have the quantity $\delta$ defined by
$\delta=g_{03}g_{03}-g_{00}g_{33}$.

The tetrad field adapted to stationary observers in spacetime (i.e.,
that satisfies conditions (17) and (18)) is written as

\begin{equation}
e_{a\mu}(t,r,\theta,\phi)=\pmatrix{-A&0&0&-C\cr
0&\sqrt{g_{11}} \,\sin\theta \cos\phi&
\sqrt{g_{22}} \cos\theta \cos\phi & -D\, r \sin\theta \sin\phi\cr
0& \sqrt{g_{11}}\, \sin\theta \sin\phi&
\sqrt{g_{22}} \cos\theta \sin\phi &  D\, r \sin\theta \cos\phi\cr
0& \sqrt{g_{11}}\, \cos\theta & -\sqrt{g_{22}}\sin\theta&0}\,.
\label{36}
\end{equation}
The functions $A, C$ and $D$ are quite similar to the corresponding
quantities in eq. (20),

\begin{eqnarray}
A(r,\theta)&=& (-g_{00})^{1/2}\,, \nonumber \\
C(r,\theta)&=&-{{g_{03}}\over{(-g_{00})^{1/2}}}\,, \nonumber \\
D(r,\theta)
&=&\biggl[ {{-\delta} \over {(r^2\sin^2\theta) g_{00}}}
\biggr]^{1/2} \,.
\label{37}
\end{eqnarray}
The same remarks about the fixation of the tetrad field given by (19)
hold in the present case.

The evaluation of $M^{ab}$ is carried out exactly as in the previous
subsection, and for this reason we will not present the details of 
the calculations. The expression below for $M^{(1)(2)}$ does not 
immediately arise as a total divergence. A number of simple 
manipulations has 
to be made in order to obtain its final form. We find that the two 
nonvanishing components of $M^{ab}$ are

\begin{equation}
M^{(1)(2)}= 2k\biggl[
\partial_1\biggl(
{{g_{03}\sqrt{g_{22}}\,\sin\theta}\over \sqrt{-g_{00}}}\biggr)+
\partial_2 \biggl(
{{g_{03}\sqrt{g_{11}}\,\cos\theta}\over \sqrt{-g_{00}}}\biggr)
\biggr]\,,
\label{38}
\end{equation}
and

\begin{eqnarray}
M^{(0)(3)}&=& 2k\biggl[
\partial_1
\biggl(
{{\delta^{1/2}\sqrt{g_{22}}\,\cos\theta}
\over {\sqrt{-g_{00}} }}
\biggr)
-\partial_2
\biggl(
{{
\delta^{1/2}\sqrt{g_{11}}\,\sin\theta
}\over {
\sqrt{-g_{00}}
}}
\biggr) \biggr]\,.
\label{39}
\end{eqnarray}
It is not difficult to verify that if we reduce the metric
components to the values of eq. (14) we arrive at eqs. (28) 
and (33). 

In order to calculate the total gravitational angular momentum we
transform definition (11) into a surface integral such that the 
surface of integration $S$, determined by the condition $r=$ 
constant, is located at spacelike infinity. Therefore we have

\begin{eqnarray}
L^{(1)(2)}
=-\int d^3x\,M^{(1)(2)}= -2k
\oint_{S\rightarrow \infty}\,d\theta d\phi
\biggl(
{{g_{03}\sqrt{g_{22}}\,\sin\theta}\over \sqrt{-g_{00}}}\biggr)\,,
\label{40}
\end{eqnarray}
assuming that the components of the metric tensor are regular at
the origin, and that $g_{03}=0$ for $r=0$.

At this point we may evaluate whether for a given spacetime metric 
tensor the total gravitational angular momentum is finite, vanishes 
or diverges. If $g_{\mu\nu}$ is given in spherical coordinates and 
if the following asymptotic behaviour is verified,

\begin{eqnarray}
g_{03}& \cong & O(1/r)+\cdots \nonumber \\
g_{22}& \cong & r^2 + O(r)+ \cdots \nonumber \\
-g_{00}& \cong & 1 + O(1/r) + \cdots\,,
\label{41}
\end{eqnarray}
expression (40) will be finite. This result agrees with the analysis
of Beig and \'O Murchadha \cite{BM} and of Szabados \cite{Sza}.

The quantity $M^{(0)(3)}$ is interpreted as the gravitational center
of mass moment. It vanished for the model described by (14) because
the latter is spherically symmetric in the limit 
$\Omega \rightarrow \infty$, and a nonvanishing $\Omega(r)$ given by
(15) and (16) does not alter the gravitational center of mass along 
the $z$ axis. The model determined by (36) is arbitrary in the sense
that the metric tensor depends on $\theta$. In view of the axial 
symmetry of the model, it is natural that the gravitational center of 
mass vanishes along the $x$ and $y$ directions. Because of the 
$\theta$ dependence of the metric tensor, (39) does not vanish
a priori.

Finaly we mention that the angular momentum of the Kerr spacetime 
will be addressed elsewhere. In the spacetime of a rotating black 
hole there exists the ergosphere, and inside the ergosphere it is not
possible to define stationary reference frames. The tetrad field 
determined by conditions (17) and (18) and that yields the Kerr
spacetime is not (mathematically) defined inside the ergosphere.

\section{The frame dependence of the gravitational angular momentum}

Definition (11) for the gravitational angular momentum is not 
invariant under local Lorentz transformations, and therefore it is
frame dependent. In this section we will discuss a physical 
aspect of this feature. Before addressing this issue, let us 
consider the frame dependence of the gravitational energy-momentum.
For this purpose we consider a black hole of mass $m$ and an
observer that is very distant from the black hole. The black hole
will appear to this observer as a particle of mass $m$, with energy
$mc^2$ ($m$ is the 
rest mass of the black hole, i.e., the mass of the black hole
in the frame where the black hole is at rest).
If, however, the black hole is
moving at velocity $v$ with respect to the observer, then its total
gravitational energy will be $\gamma m c^2$, where 
$\gamma=(1-v^2/c^2)^{-1/2}$. This example is a consequence of
the special theory of relativity, and demonstrates the 
frame dependence of the gravitational energy-momentum. We note that
the frame dependence is not restricted to observers at spacelike 
infinity. It holds for observers located
everywhere in the three-dimensional space.

The dependence of the gravitational angular momentum on the frame is
an intrinsic feature of definition (11). In order to understand some 
consequences of this 
feature, we will investigate a special set of tetrad fields. Let us 
consider the frame for the spacetime of the rotating neutron star that 
satisfies Schwinger's time gauge condition,

\begin{equation}
e_{(i)}\,^0=0\,.
\label{42}
\end{equation}
It reads

\begin{equation}
e_{a\mu}=\pmatrix{-\alpha&0&0&0\cr
\Omega r\,\sin\theta\,\sin\phi&\beta\,\sin\theta\,\cos\phi&
r\,\cos\theta\,\cos\phi&-r\,\sin\theta\,\sin\phi\cr
-\Omega r\,\sin\theta\,\cos\phi&\beta\,\sin\theta\,\sin\phi&
r\,\cos\theta\,\sin\phi&r\,\sin\theta\,\cos\phi\cr
0&\beta\,\cos\theta&-r\,\sin\theta&0}\,,
\label{43}
\end{equation}
where $\alpha$, $\beta$ and $\Omega$ are the quantities defined in
(14-16). This frame is adapted to the field of observers whose
velocity $e_{(0)}\,^\mu$ is given by

\begin{equation}
e_{(0)}\,^\mu(t,r,\theta,\phi)=
{1\over \alpha}(1,0,0,\Omega(r)\,)\,.
\label{44}
\end{equation}
$\Omega(r)$ is the dragging velocity of inertial frames that rotate 
under the action of the neutron star.
The expression above for $e_{(0)}\,^\mu$ describes the velocity
field of observers in circular motion around the star. 
It turns out that the angular momentum of the gravitational 
field calculated out of (43) vanishes:
$L^{(1)(2)}=0$. It is not difficult to arrive at this result. Out of
eq. (43) we find

\begin{eqnarray}
T_{013}&=& -(1-\beta)\Omega r\sin^2\theta \,, \nonumber \\
T_{212}&=& (1-\beta)r \,, \nonumber \\
T_{313}&=& (1-\beta)r\sin^2\theta \,. 
\label{45}
\end{eqnarray}
Working out and simplifying the expression for $M^{(1)(2)}$ we obtain

\begin{eqnarray}
M^{(1)(2)}&=& -2ke\biggl[
(e^{(1)}\,_0e^{(2)}\,_1-e^{(1)}\,_1e^{(2)}\,_0)g^{11}[(g^{00}g^{33}
-g^{03}g^{03})T_{313}\nonumber \\
&{}&+g^{00}g^{22}T_{212}] \nonumber \\
&{}&+(e^{(1)}\,_1e^{(2)}\,_3-e^{(1)}\,_3e^{(2)}\,_1)g^{11}
[(g^{00}g^{33}-g^{03}g^{03})T_{013}\nonumber \\
&{}&-g^{03}g^{22}T_{212}]
\biggr]
\,. \label{46}
\end{eqnarray}
Taking into account

\begin{eqnarray}
(g^{00}g^{33}-g^{03}g^{03})T_{313}+g^{00}g^{22}T_{212}
&=& -\frac{2}{\delta}\,T_{212}\sin^2\theta  \,, \nonumber \\
(g^{00}g^{33}-g^{03}g^{03})T_{013}-g^{03}g^{22}T_{212}
&=&\frac{2}{\delta}\,\Omega\,
T_{212}\sin^2\theta  \,, 
\label{47}
\end{eqnarray}
and also that

\begin{eqnarray}
e^{(1)}\,_0e^{(2)}\,_1-e^{(1)}\,_1e^{(2)}\,_0
&=&  \Omega \beta r\sin^2\theta \,, \nonumber \\
e^{(1)}\,_1e^{(2)}\,_3-e^{(1)}\,_3e^{(2)}\,_1
&=&\beta r\sin^2\theta  \,,
\label{48}
\end{eqnarray}
we find that $M^{(1)(2)}=0$. The other components also vanish. We
note that observers with four-velocity given by (44) are
known in the literature as ZAMOs (zero angular momentum observers)
(see Ref. \cite{Schutz}, section 11.3). These observers follow
trajectories with constant radial coordinate $r$ and with angular 
velocity given by the dragging velocity of inertial frames.

The result above shows that 
observers that are in rotational motion around the rotating source
measure the gravitational angular momentum differently from 
static observers. An explanation for this result must take into 
account the angular momentum of the source, which is different for
observers at rest and for those that rotate around the source.
In the Newtonian limit of the theory the angular momentum of the 
source in the frame of observers that rotate at the same angular 
frequency vanishes. We know that this feature holds for a rigid body
in Newtonian mechanics, where the angular momentum depends not only 
on the frame, but also on the origin of the frame.

We conclude that definition (11) for the angular momentum of 
the field is a measure of the dragging of inertial frames. 
Observers whose angular velocity around the rotating
source is the same as the dragging velocity $\Omega(r)$ do not 
measure the dragging velocity itself (and possibly other dragging 
effects), and therefore for these observers the 
gravitational angular momentum vanishes. 

Because of the frame dependence, the interpretation of definition
(11) is different from the standard interpretation of the 
gravitational angular momentum \cite{BM,Sza} based on the idea of 
conserved field quantities. The latter are obtained by means of 
surface integrals at spacelike infinity. In our opinion these are 
just two different interpretations that do not seem to be in 
conflict.

\section{Conclusions}

The major reason for considering the present definition of 
gravitational angular momentum is that it satisfies the Poincar\'e 
algebra. In the framework of the TEGR the energy-momentum and 
angular momemtum of the gravitational field are defined by suitably 
interpreting the constraint equations of the theory, not by means of
boundary integrals of the Hamiltonian. Definitions (6) and (11) are
evaluated in the configuration space of the theory.

The definition for the gravitational angular momentum investigated
here yields consistent results in the consideration of 
a rotating neutron star. 
In the context of isolated rotating sources, we observe that
the angular momentum of the gravitational field is related to the
dragging velocity of inertial frames $\Omega(r)$. 
By inspecting eqs. (28) and
(38) we see that the angular momentum density depends on $g_{03}$, 
and typically we have $g_{03}\cong -\Omega(r) r^2\sin^2\theta$,
at least at great distances from the source. Therefore the
angular momentum of the spacetime is a closely related to the 
dragging of inertial frames.

One main result of the paper 
is eq. (40), which gives the gravitational angular momentum
generated by an isolated rotating source. Equation
(40) allows to verify precisely which is the asymptotic behaviour of
the metric tensor that leads to a finite value for the angular 
momentum in the frame of stationary observers. In the consideration 
of a rotating neutron star, we found that if $J$ is the angular 
momentum of the star, the angular momentum of the gravitational field
is $2J/3$, in the limit of rigid rotation. 
So far we do not have an explanation as to why the angular 
momentum of the field is $2J/3$ (and not $J$).
This issue will be addressed in the
future. However, to our knowledge there is no physical requirement 
for the two angular momenta, of the field and of the source, to be 
the same.

\bigskip
\noindent {\bf Acknowledgement}\par
\noindent This work was supported in part by CNPQ.

\end{document}